\documentclass[10pt, a4paper, twocolumn]{article} 

\usepackage[english]{babel} 

\usepackage{microtype} 

\usepackage{amsmath,amsfonts,amsthm} 

\usepackage[svgnames]{xcolor} 

\usepackage[hang, small, labelfont=bf, up, textfont=it]{caption} 

\usepackage{booktabs} 

\usepackage{lastpage} 

\usepackage{graphicx} 
\usepackage{subcaption}
\usepackage[title]{appendix}
\usepackage{amsfonts}
\usepackage{breqn}
\DeclareUnicodeCharacter{2212}{-}
\usepackage{enumitem} 
\setlist{noitemsep} 

\usepackage{sectsty} 
\allsectionsfont{\usefont{OT1}{phv}{b}{n}} 

\usepackage{geometry} 

\usepackage[T1]{fontenc} 
\usepackage[utf8]{inputenc} 

\usepackage{XCharter} 

\usepackage{fancyhdr} 
\fancypagestyle{firstpage}{ 
	\fancyhf{}
	\chead{©2020 The Johns Hopkins University Applied Physics Laboratory LLC}
}

\newcommand{\authorstyle}[1]{{\large\usefont{OT1}{phv}{b}{n}\color{DarkRed}#1}} 

\newcommand{\institution}[1]{{\footnotesize\usefont{OT1}{phv}{m}{sl}\color{Black}#1}} 

\usepackage{titling} 

\newcommand{\HorRule}{\color{DarkGoldenrod}\rule{\linewidth}{1pt}} 

\pretitle{
	\vspace{-30pt} 
	\HorRule\vspace{10pt} 
	\fontsize{32}{36}\usefont{OT1}{phv}{b}{n}\selectfont 
	\color{DarkRed} 
}

\posttitle{\par\vskip 15pt} 

\preauthor{} 

\postauthor{ 
	\vspace{10pt} 
	\par\HorRule 
	\vspace{20pt} 
}

\usepackage{lettrine} 
\usepackage{fix-cm}	

\newcommand{\initial}[1]{ 
	\lettrine[lines=3,findent=4pt,nindent=0pt]{
		\color{DarkGoldenrod}
		{#1}
	}{}%
}

\usepackage{xstring} 

\newcommand{\lettrineabstract}[1]{
	\StrLeft{#1}{1}[\firstletter] 
	\initial{\firstletter}\textbf{\StrGobbleLeft{#1}{1}} 
}

\usepackage[backend=bibtex,natbib=true]{biblatex}
\usepackage[autostyle=true]{csquotes} 

\title{RF PIX2PIX Unsupervised Wi-Fi to Video Translation} 

\author{
	\authorstyle{Michael Drob\textsuperscript{1}} 
	\newline\newline 
	\textsuperscript{1}\institution{Johns Hopkins Applied Physics Lab, Maryland, United States of America}\\ 
}

\date{\today} 

\begin{document}

\maketitle 

\thispagestyle{firstpage} 


\begin{figure}[h]
\centering
    \includegraphics[width=\linewidth]{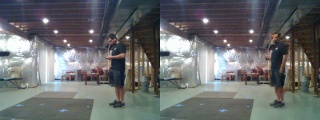}
    	\begin{minipage}[b]{.5\linewidth}
	\centering
	\textnormal{Webcam}
	\end{minipage}%
	\begin{minipage}[b]{.5\linewidth}
	\centering
	\textnormal{From Wi-Fi scatter}
	\end{minipage}
    \caption{Webcam vs Video recovered from Wi-Fi scatter}
    \label{fig:Intro}
\end{figure}

\lettrineabstract{With the proliferation of Wi-Fi devices in the environment, our surroundings are increasingly illuminated with low level RF scatter. This scatter illuminates objects in the environment much like radar or LIDAR. We show that a novel unsupervised network, based on the PIX2PIX GAN architecture \cite{isola2018imagetoimage}, can recover and visually reconstruct scene information solely from Wi-Fi background energy; in contrast to a significantly less accurate approach by Kefayati (et. all) \cite{2020arXiv200105842H} which requires careful object labeling to recover object location from a scene. This is accomplished by learning a more robust mapping function between the channel state information (CSI) from Wi-Fi packets and Video image sample distributions. }

\section{Introduction}
Within the visible spectrum, light sources emit electromagnetic (EM) radiation that scatters across objects in the scene eventually reaching our eye where its reconstructed as images. Similarly, EM radiation from sources outside the visible spectrum, like Wi-Fi access points (WAP) scatters across objects in the environment. Until recently reconstructing images from this scattered energy was intractable due to complexities and ambiguities in the information received. We present a novel unsupervised approach to developing a  robust mapping between the visible and invisible electromagnetic spectrum.  

Intersection over Union (IoU) is a common method for evaluating the quality of bounding boxes. It is a ratio of the overlap between a bounding box and ground truth divided by the union of both boxes. A ratio of .50 for this metric (IoU-50) is often used as minimum threshold to determine the accuracy of a bounding box\ref{eqn:IoU}. With one person in the scene we calculate an average precision intersection over union thershold 50 (AP IoU-50) score of 92\%. This is an 18\% improvement over the accuracy we were able to obtain with a supervised approach on the same input data. 

\begin{equation}
	IoU = 
	\dfrac {Overlap_{boxes}}{Union_{boxes}} 
\label{eqn:IoU}
\end{equation}

\section{Setup}

We positioned a Wireless AP and a laptop 28' apart. Using 2 Ettus N210 recievers, synced in MIMO mode, and GNURadio software we captured Wi-Fi packets from the WAP, filtering on media access control address (MAC) address. Video frames from the laptops webcam were captured syncronosly with the reception of Wi-Fi packets.  

\begin{figure}[h]
\centering
  \begin{subfigure}[h]{0.2\textwidth}
    \includegraphics[width=\textwidth]{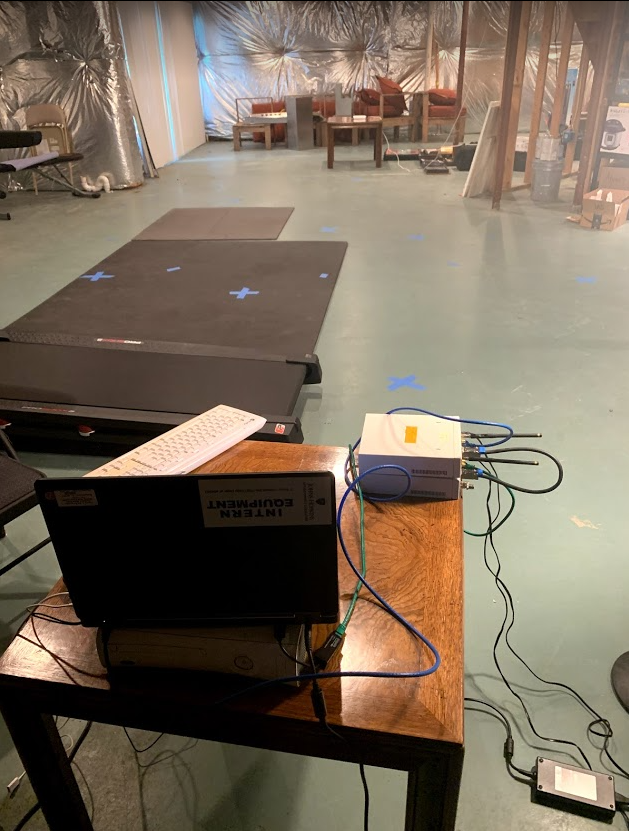}
    \subcaption{Ettus MIMO N210's} 
    \label{fig:1}
  \end{subfigure}
  \begin{subfigure}[h]{0.2\textwidth}
    \includegraphics[width=\textwidth]{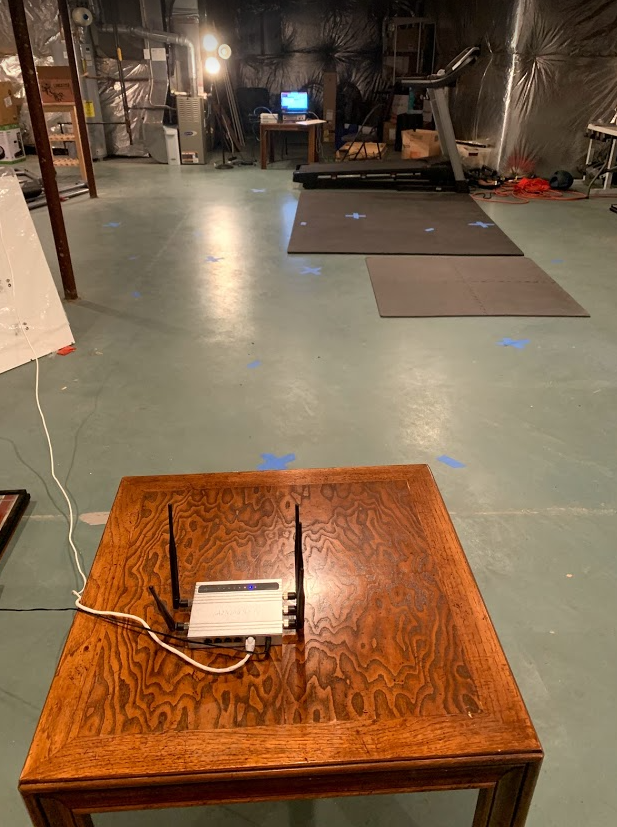}
    \subcaption{Wireless AP}
    \label{fig:2}
  \end{subfigure}
  \caption{Test Environment}
  \label{Test Environment}
\end{figure}


\subsection{CSI Calculation}
The channel state information (CSI) indicates the change that the signal experiences while traversing the channel \cite{9149443}. Its highly dependent on multipath and scattering caused by objects in the scene. Because of this we use it as a basis for transfering between the RF and visible light domains.

Using GNURadio we calculated the Channel State Information (CSI) vectors for each Antenna. For the case of 20MHz bandwidth Wi-Fi we get one complex value for each of 56 subcarrier(OFDM). The complex and real portion of each value is channelized resulting in a CSI tensor dimension of [4, 1, 56] \ref{eqn:csi}.

\begin{equation}
	CSI = 
	\begin{bmatrix}
		I_{sc1} & I_{sc2} & .. &  I_{sc56}  \\
		Q_{sc1} & Q_{sc2} & .. &  Q_{sc56} \\
		I_{sc1} & I_{sc2} & .. &  I_{sc56}  \\
		Q_{sc1} & Q_{sc2} & .. &  Q_{sc56} 
	\end{bmatrix}
\label{eqn:csi}
\end{equation}


\subsection{Normalization}

For normalization we performed regular min/max to bring the CSI input values within a range of 0-1 \ref{eqn:Normalization}. We did not see any benefit in performing phase sanitization as described in  \cite{2020arXiv200105842H,Keerativoranan_2018,Qian}.

\begin{equation}
	\textsuperscript{$\wedge$}CSI = 
	\dfrac {CSI - min(CSI)}{max(CSI)-min(CSI)} 
\label{eqn:Normalization}
\end{equation}

\section{Supervised Training}

Much like Kefayati (et. all) \cite{2020arXiv200105842H} we began by training a network to identify dynamic objects of interest in the scene. We did this by identifying objects using Torchvisions Faster R-CNN ResNet50 \cite{NIPS2015_14bfa6bb} model pre-trained on COCO\cite{lin2015microsoft}. Detections from this model with > 90\% confidence are used as labels for dynamic objects in the scene. Our inference network uses ResNet18 also mapped to output prediction boxes with the COCO coordinates. The loss function for training becomes equation \ref{eqn:loss}. 

\begin{equation}
	Loss = 
	MSE( RCNN(Video)_{c>90}, ResNet18(CSI))
\label{eqn:loss}
\end{equation}

Results from the training procedure are illustrated in Appendix A with a calculated AP IOU50 of 74\%.

\section{Unsupervised Training}

The task of converting CSI to video can be summerized as functional mapping between two domains\ref{eqn:map}. In the unsupervised case we would like to enforce a consistant mapping. Where for any 2 time pairs t0 and t1 the set \[CSI_{t0}, G(CSI_{t0})\] and \[CSI_{t1}, Video_{t1}\] are indistingushible by a descriminator function D.  This is expressed by the GAN objective \ref{eqn:objective}\cite{isola2018imagetoimage}. To satisfy this objective the function G must find a true mapping between CSI and video given to it by the environment. This technique has shown state of the art results transforming images in other domains \ref{fig:shoe1}\cite{isola2018imagetoimage}.

Results from the training procedure are illustrated in Appendix B.

\begin{equation}
	Video = G(CSI)
\label{eqn:map}
\end{equation}
\begin{dmath}
 \mathcal{L}_{cGAN}\left(G, D\right) =\mathbb{E}_{x,y}\left[\log D\left(x, y\right)\right]+
\mathbb{E}_{x,z}\left[log(1 − D\left(x, G\left(x, z\right)\right)\right] 
\label{eqn:objective}
\end{dmath}

\begin{figure}[h]
\centering
    \includegraphics{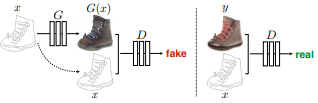}
    \caption{Image-to-Image Translation with Conditional Adversarial Networks} 
    \label{fig:shoe1}
\end{figure}

\subsection{PatchGAN Descriminator Modification}
\begin{figure}[h]
\centering
    \includegraphics[width=\linewidth]{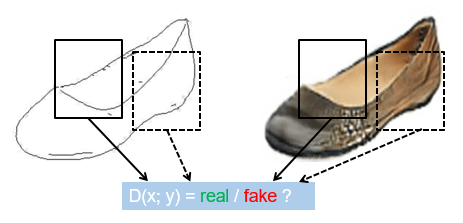}
    \caption{Original PatchGAN Discriminator} 
    \label{fig:shoepatch}
\end{figure}

In the original PatchGAN \cite{isola2018imagetoimage} paper the discriminator can classify whether 70×70 overlapping patches are real or fake \ref{fig:shoepatch}. Each patch consists of 6 input channels, RGB for the input image x and RGB for the output image y, which focus on a corresponding sub region of each pair. This allows for a smaller discriminator network with simpler training and a better focus on image detail. 

\begin{figure}[h]
\centering
    \includegraphics[width=\linewidth]{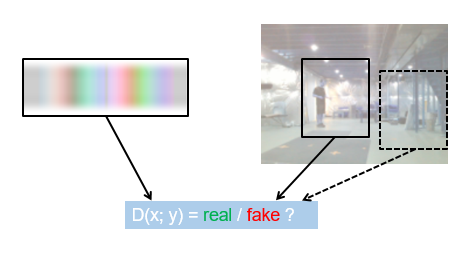}
    \caption{Modified PatchGAN Discriminator- Every patch from the output on the right is paired with the full CSI input visualized on the left} 
    \label{fig:csi}
\end{figure}

In the case of CSI to video translation the original PatchGAN discriminator must be modified for successful training. One issue is that the input CSI data’s shape [4,1,56] did not match the output image y resolution shape of [3,160,120]. Additionally, we recognized that each pixel in the output image space depends on all the input CSI information. This is a result of each object in the scene contributing to the multipath wireless environment. To better match this characteristic of the data we designed a new PatchGAN that still used 70x70 overlapping patches on the output data y and paired it with the full CSI data input x for each real fake classification \ref{fig:csi}. 

\subsection{Unsupervised Training Results}

\begin{figure}[h]
\centering
    \includegraphics[width=\linewidth]{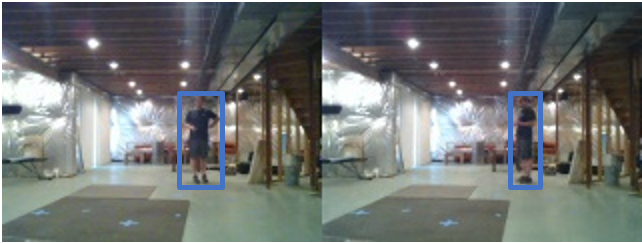}
    	\begin{minipage}[b]{.5\linewidth}
	\centering
	\textnormal{Webcam}
	\end{minipage}%
	\begin{minipage}[b]{.5\linewidth}
	\centering
	\textnormal{From Wi-Fi scatter}
	\end{minipage}
    \caption{Webcam vs Video recovered from Wi-Fi CSI with bounding boxes}
    \label{fig:results}
\end{figure}

In order to compare the results of the unsupervised training model vs that of the supervised, we devised a method for calculating an AP IoU-50 score for the CSI output images. Just like we did for the supervised case we used Torchvision’s pre-trained RCNN ResNet50 model to label objects within the image captures from the webcam. For the images created from the CSI data we hand annotated the labels \ref{fig:results}. This was done because the CSI generated images could be missing some of the finer details the RCNN model uses to classify people. The bounding boxes are then compared in the standard AP IoU-50 fashion resulting in a score of 92\%.

\subsection{Results Summary}

Electromagnetic radiation from visible light sources or Wi-Fi access points, interact similarly with the environment, scattering and refracting off objects on the way to a receiver. We showed that mapping input data from the invisible Wi-Fi domain to the visible domain captured by webcams can be accomplished in a unsupervised way. We developed a novel modification to a state of the art domain transfer unsupervised generative model Pix2Pix \cite{isola2018imagetoimage}. This methodology results in a more robust mapping with a measured 18\% improvement in accuracy vs previously used supervised methods. 

\subsection{Next Steps}

The results we show in the this paper show excellent performance in reconstructing video in a simple environment. In our scene we only had one person as a object of interest with consistent motion. We did have other motion outside of the vision of the webcam that, encouragingly, showed little effect on the quality of the mapping. We also experimented with 2 people in the scene with positive results. More work needs to be done evaluating this performance before publication. 

\newpage

\onecolumn{
	\printbibliography[title={Bibliography}] 
}

\newpage

\onecolumn
\centering
\begin{appendices}
\section{Supervised Training Results}

\captionsetup[subfigure]{font=bf}
\begin{figure}
\centering
        	\begin{minipage}[b]{.3\linewidth}
	\centering
	\subcaption{Capture1}
	\end{minipage}%
	\begin{minipage}[b]{.3\linewidth}
	\centering
	\subcaption{Capture2}
	\end{minipage}
	\begin{minipage}[b]{.3\linewidth}
	\centering
	\subcaption{Capture3}
	\end{minipage}

  \begin{subfigure}[b]{0.3\textwidth}
    \includegraphics[width=\textwidth]{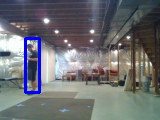}
    \label{fig:1}
  \end{subfigure}
  \begin{subfigure}[b]{0.3\textwidth}
    \includegraphics[width=\textwidth]{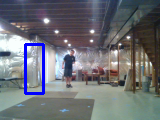}
    \label{fig:2}
  \end{subfigure}
  \begin{subfigure}[b]{0.3\textwidth}
    \includegraphics[width=\textwidth]{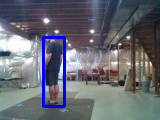}
    \label{fig:2}
  \end{subfigure}

        	\begin{minipage}[b]{.3\linewidth}
	\centering
	\subcaption{Capture4}
	\end{minipage}%
	\begin{minipage}[b]{.3\linewidth}
	\centering
	\subcaption{Capture5}
	\end{minipage}
	\begin{minipage}[b]{.3\linewidth}
	\centering
	\subcaption{Capture6}
	\end{minipage}

  \begin{subfigure}[b]{0.3\textwidth}
    \includegraphics[width=\textwidth]{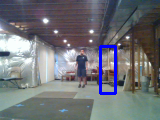}
    \label{fig:1}
  \end{subfigure}
  \begin{subfigure}[b]{0.3\textwidth}
    \includegraphics[width=\textwidth]{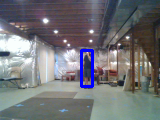}
    \label{fig:2}
  \end{subfigure}
  \begin{subfigure}[b]{0.3\textwidth}
    \includegraphics[width=\textwidth]{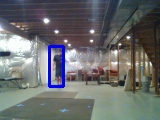}
    \label{fig:2}
  \end{subfigure}

        	\begin{minipage}[b]{.3\linewidth}
	\centering
	\subcaption{Capture7}
	\end{minipage}%
	\begin{minipage}[b]{.3\linewidth}
	\centering
	\subcaption{Capture8}
	\end{minipage}
	\begin{minipage}[b]{.3\linewidth}
	\centering
	\subcaption{Capture9}
	\end{minipage}

  \begin{subfigure}[b]{0.3\textwidth}
    \includegraphics[width=\textwidth]{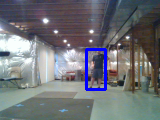}
    \label{fig:1}
  \end{subfigure}
  \begin{subfigure}[b]{0.3\textwidth}
    \includegraphics[width=\textwidth]{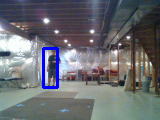}
    \label{fig:2}
  \end{subfigure}
  \begin{subfigure}[b]{0.3\textwidth}
    \includegraphics[width=\textwidth]{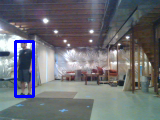}
    \label{fig:2}
  \end{subfigure}

        	\begin{minipage}[b]{.3\linewidth}
	\centering
	\subcaption{Capture10}
	\end{minipage}%
	\begin{minipage}[b]{.3\linewidth}
	\centering
	\subcaption{Capture11}
	\end{minipage}
	\begin{minipage}[b]{.3\linewidth}
	\centering
	\subcaption{Capture12}
	\end{minipage}

  \begin{subfigure}[b]{0.3\textwidth}
    \includegraphics[width=\textwidth]{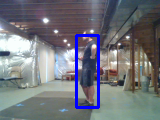}
    \label{fig:1}
  \end{subfigure}
  \begin{subfigure}[b]{0.3\textwidth}
    \includegraphics[width=\textwidth]{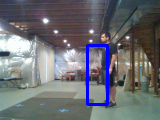}
    \label{fig:2}
  \end{subfigure}
  \begin{subfigure}[b]{0.3\textwidth}
    \includegraphics[width=\textwidth]{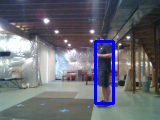}
    \label{fig:2}
  \end{subfigure}

        	\begin{minipage}[b]{.3\linewidth}
	\centering
	\subcaption{Capture13}
	\end{minipage}%
	\begin{minipage}[b]{.3\linewidth}
	\centering
	\subcaption{Capture14}
	\end{minipage}
	\begin{minipage}[b]{.3\linewidth}
	\centering
	\subcaption{Capture15}
	\end{minipage}

  \begin{subfigure}[b]{0.3\textwidth}
    \includegraphics[width=\textwidth]{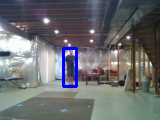}
    \label{fig:1}
  \end{subfigure}
  \begin{subfigure}[b]{0.3\textwidth}
    \includegraphics[width=\textwidth]{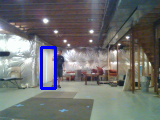}
    \label{fig:2}
  \end{subfigure}
  \begin{subfigure}[b]{0.3\textwidth}
    \includegraphics[width=\textwidth]{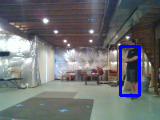}
    \label{fig:2}
  \end{subfigure}
\caption{Bounding Boxes Using Wi-Fi Input}
\end{figure}

\newpage

\centering
\section{Unsupervised Training Results}

\captionsetup[subfigure]{font=bf}
\begin{figure}
        	\begin{minipage}[b]{.5\linewidth}
	\centering
	\subcaption{Capture1}
	\end{minipage}%
	\begin{minipage}[b]{.5\linewidth}
	\centering
	\subcaption{Capture2}
	\end{minipage}
  \begin{subfigure}[b]{0.5\textwidth}
    \includegraphics[width=\textwidth]{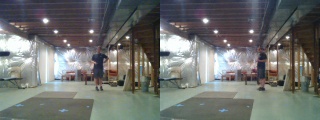}
    	\begin{minipage}[b]{.5\linewidth}
	\centering
	\textnormal{Webcam}
	\end{minipage}%
	\begin{minipage}[b]{.5\linewidth}
	\centering
	\textnormal{From Wi-Fi}
	\end{minipage}
    \label{fig:1}
  \end{subfigure}
  \begin{subfigure}[b]{0.5\textwidth}
    \includegraphics[width=\textwidth]{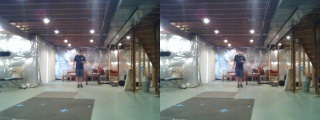}
    	\begin{minipage}[b]{.5\linewidth}
	\centering
	\textnormal{Webcam}
	\end{minipage}%
	\begin{minipage}[b]{.5\linewidth}
	\centering
	\textnormal{From Wi-Fi}
	\end{minipage}
    \label{fig:2}
  \end{subfigure}
        	\begin{minipage}[b]{.5\linewidth}
	\centering
	\subcaption{Capture3}
	\end{minipage}%
	\begin{minipage}[b]{.5\linewidth}
	\centering
	\subcaption{Capture4}
	\end{minipage}
  \begin{subfigure}[b]{0.5\textwidth}
    \includegraphics[width=\textwidth]{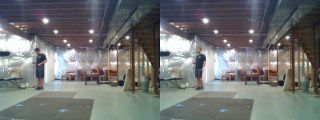}
    	\begin{minipage}[b]{.5\linewidth}
	\centering
	\textnormal{Webcam}
	\end{minipage}%
	\begin{minipage}[b]{.5\linewidth}
	\centering
	\textnormal{From Wi-Fi}
	\end{minipage}
    \label{fig:1}
  \end{subfigure}
  \begin{subfigure}[b]{0.5\textwidth}
    \includegraphics[width=\textwidth]{./val/pix2pix3.jpg}
    	\begin{minipage}[b]{.5\linewidth}
	\centering
	\textnormal{Webcam}
	\end{minipage}%
	\begin{minipage}[b]{.5\linewidth}
	\centering
	\textnormal{From Wi-Fi}
	\end{minipage}
    \label{fig:2}
  \end{subfigure}
        	\begin{minipage}[b]{.5\linewidth}
	\centering
	\subcaption{Capture5}
	\end{minipage}%
	\begin{minipage}[b]{.5\linewidth}
	\centering
	\subcaption{Capture6}
	\end{minipage}
  \begin{subfigure}[b]{0.5\textwidth}
    \includegraphics[width=\textwidth]{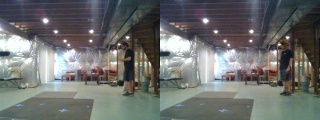}
    	\begin{minipage}[b]{.5\linewidth}
	\centering
	\textnormal{Webcam}
	\end{minipage}%
	\begin{minipage}[b]{.5\linewidth}
	\centering
	\textnormal{From Wi-Fi}
	\end{minipage}
    \label{fig:1}
  \end{subfigure}
  \begin{subfigure}[b]{0.5\textwidth}
    \includegraphics[width=\textwidth]{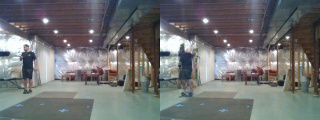}
    	\begin{minipage}[b]{.5\linewidth}
	\centering
	\textnormal{Webcam}
	\end{minipage}%
	\begin{minipage}[b]{.5\linewidth}
	\centering
	\textnormal{From Wi-Fi}
	\end{minipage}
    \label{fig:2}
  \end{subfigure}
        	\begin{minipage}[b]{.5\linewidth}
	\centering
	\subcaption{Capture7}
	\end{minipage}%
	\begin{minipage}[b]{.5\linewidth}
	\centering
	\subcaption{Capture8}
	\end{minipage}
  \begin{subfigure}[b]{0.5\textwidth}
    \includegraphics[width=\textwidth]{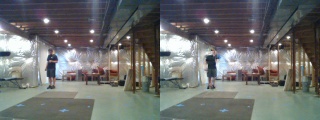}
    	\begin{minipage}[b]{.5\linewidth}
	\centering
	\textnormal{Webcam}
	\end{minipage}%
	\begin{minipage}[b]{.5\linewidth}
	\centering
	\textnormal{From Wi-Fi}
	\end{minipage}
    \label{fig:1}
  \end{subfigure}
  \begin{subfigure}[b]{0.5\textwidth}
    \includegraphics[width=\textwidth]{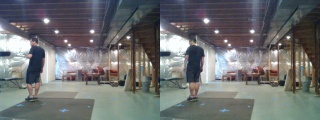}
    	\begin{minipage}[b]{.5\linewidth}
	\centering
	\textnormal{Webcam}
	\end{minipage}%
	\begin{minipage}[b]{.5\linewidth}
	\centering
	\textnormal{From Wi-Fi}
	\end{minipage}
    \label{fig:2}
  \end{subfigure}
        	\begin{minipage}[b]{.5\linewidth}
	\centering
	\subcaption{Capture9}
	\end{minipage}%
	\begin{minipage}[b]{.5\linewidth}
	\centering
	\subcaption{Capture10}
	\end{minipage}
  \begin{subfigure}[b]{0.5\textwidth}
    \includegraphics[width=\textwidth]{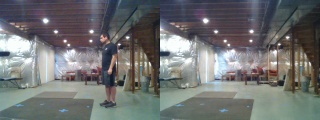}
    	\begin{minipage}[b]{.5\linewidth}
	\centering
	\textnormal{Webcam}
	\end{minipage}%
	\begin{minipage}[b]{.5\linewidth}
	\centering
	\textnormal{From Wi-Fi}
	\end{minipage}
    \label{fig:1}
  \end{subfigure}
  \begin{subfigure}[b]{0.5\textwidth}
    \includegraphics[width=\textwidth]{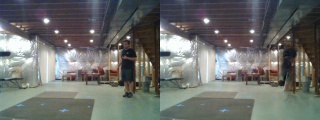}
    	\begin{minipage}[b]{.5\linewidth}
	\centering
	\textnormal{Webcam}
	\end{minipage}%
	\begin{minipage}[b]{.5\linewidth}
	\centering
	\textnormal{From Wi-Fi}
	\end{minipage}
    \label{fig:2}
  \end{subfigure}
\caption{Unsupervised CSI to Video Translation}
\end{figure}

\captionsetup[subfigure]{font=bf}
\begin{figure}
        	\begin{minipage}[b]{.5\linewidth}
	\centering
	\subcaption{Capture11}
	\end{minipage}%
	\begin{minipage}[b]{.5\linewidth}
	\centering
	\subcaption{Capture12}
	\end{minipage}
  \begin{subfigure}[b]{0.5\textwidth}
    \includegraphics[width=\textwidth]{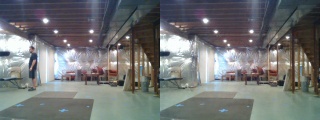}
    	\begin{minipage}[b]{.5\linewidth}
	\centering
	\textnormal{Webcam}
	\end{minipage}%
	\begin{minipage}[b]{.5\linewidth}
	\centering
	\textnormal{From Wi-Fi}
	\end{minipage}
    \label{fig:1}
  \end{subfigure}
  \begin{subfigure}[b]{0.5\textwidth}
    \includegraphics[width=\textwidth]{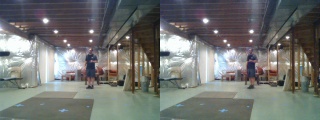}
    	\begin{minipage}[b]{.5\linewidth}
	\centering
	\textnormal{Webcam}
	\end{minipage}%
	\begin{minipage}[b]{.5\linewidth}
	\centering
	\textnormal{From Wi-Fi}
	\end{minipage}
    \label{fig:2}
  \end{subfigure}
        	\begin{minipage}[b]{.5\linewidth}
	\centering
	\subcaption{Capture13}
	\end{minipage}%
	\begin{minipage}[b]{.5\linewidth}
	\centering
	\subcaption{Capture14}
	\end{minipage}
  \begin{subfigure}[b]{0.5\textwidth}
    \includegraphics[width=\textwidth]{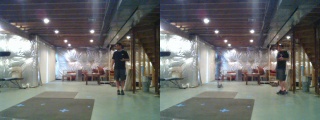}
    	\begin{minipage}[b]{.5\linewidth}
	\centering
	\textnormal{Webcam}
	\end{minipage}%
	\begin{minipage}[b]{.5\linewidth}
	\centering
	\textnormal{From Wi-Fi}
	\end{minipage}
    \label{fig:1}
  \end{subfigure}
  \begin{subfigure}[b]{0.5\textwidth}
    \includegraphics[width=\textwidth]{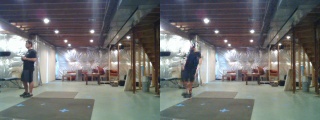}
    	\begin{minipage}[b]{.5\linewidth}
	\centering
	\textnormal{Webcam}
	\end{minipage}%
	\begin{minipage}[b]{.5\linewidth}
	\centering
	\textnormal{From Wi-Fi}
	\end{minipage}
    \label{fig:2}
  \end{subfigure}
        	\begin{minipage}[b]{.5\linewidth}
	\centering
	\subcaption{Capture15}
	\end{minipage}%
	\begin{minipage}[b]{.5\linewidth}
	\centering
	\subcaption{Capture16}
	\end{minipage}
  \begin{subfigure}[b]{0.5\textwidth}
    \includegraphics[width=\textwidth]{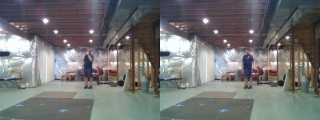}
    	\begin{minipage}[b]{.5\linewidth}
	\centering
	\textnormal{Webcam}
	\end{minipage}%
	\begin{minipage}[b]{.5\linewidth}
	\centering
	\textnormal{From Wi-Fi}
	\end{minipage}
    \label{fig:1}
  \end{subfigure}
  \begin{subfigure}[b]{0.5\textwidth}
    \includegraphics[width=\textwidth]{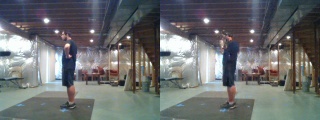}
    	\begin{minipage}[b]{.5\linewidth}
	\centering
	\textnormal{Webcam}
	\end{minipage}%
	\begin{minipage}[b]{.5\linewidth}
	\centering
	\textnormal{From Wi-Fi}
	\end{minipage}
    \label{fig:2}
  \end{subfigure}
        	\begin{minipage}[b]{.5\linewidth}
	\centering
	\subcaption{Capture17}
	\end{minipage}%
	\begin{minipage}[b]{.5\linewidth}
	\centering
	\subcaption{Capture18}
	\end{minipage}
  \begin{subfigure}[b]{0.5\textwidth}
    \includegraphics[width=\textwidth]{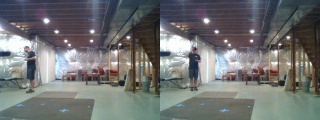}
    	\begin{minipage}[b]{.5\linewidth}
	\centering
	\textnormal{Webcam}
	\end{minipage}%
	\begin{minipage}[b]{.5\linewidth}
	\centering
	\textnormal{From Wi-Fi}
	\end{minipage}
    \label{fig:1}
  \end{subfigure}
  \begin{subfigure}[b]{0.5\textwidth}
    \includegraphics[width=\textwidth]{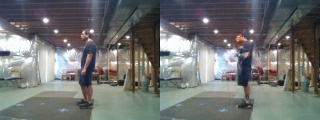}
    	\begin{minipage}[b]{.5\linewidth}
	\centering
	\textnormal{Webcam}
	\end{minipage}%
	\begin{minipage}[b]{.5\linewidth}
	\centering
	\textnormal{From Wi-Fi}
	\end{minipage}
    \label{fig:2}
  \end{subfigure}
        	\begin{minipage}[b]{.5\linewidth}
	\centering
	\subcaption{Capture19}
	\end{minipage}%
	\begin{minipage}[b]{.5\linewidth}
	\centering
	\subcaption{Capture20}
	\end{minipage}
  \begin{subfigure}[b]{0.5\textwidth}
    \includegraphics[width=\textwidth]{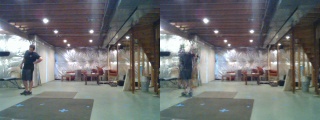}
    	\begin{minipage}[b]{.5\linewidth}
	\centering
	\textnormal{Webcam}
	\end{minipage}%
	\begin{minipage}[b]{.5\linewidth}
	\centering
	\textnormal{From Wi-Fi}
	\end{minipage}
    \label{fig:1}
  \end{subfigure}
  \begin{subfigure}[b]{0.5\textwidth}
    \includegraphics[width=\textwidth]{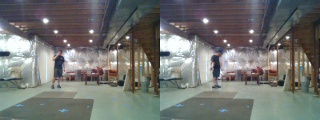}
    	\begin{minipage}[b]{.5\linewidth}
	\centering
	\textnormal{Webcam}
	\end{minipage}%
	\begin{minipage}[b]{.5\linewidth}
	\centering
	\textnormal{From Wi-Fi}
	\end{minipage}
    \label{fig:2}
  \end{subfigure}
\caption{Unsupervised CSI to Video Translation}
\end{figure}

\end{appendices}

\end{document}